\documentstyle [12pt,epsf] {article}

\catcode`@=12
\newcommand{\beq}{\begin{equation}}
\newcommand{\eeq}{\end{equation}}
\newcommand{\nn}{\nonumber}
\newcommand{\bea}{\begin{eqnarray}}
\newcommand{\eea}{\end{eqnarray}}

\newcommand{\rfn}[1]{(\ref{#1})}
\newcommand{\Eq}[1]{Eq.~(\ref{#1})}

\newcommand{\ea}{{\it et al.}}

\newcommand{\np}[1]{{ Nucl. Phys. }{\bf #1}}
\newcommand{\plet}[1]{{ Phys. Lett. }{\bf #1}}
\newcommand{\pr}[1]{{ Phys. Rev. }{\bf #1}}
\newcommand{\prlet}[1]{{ Phys. Rev. Lett. }{\bf #1}}

\def\lsim{\mathrel{\vcenter{\hbox{$<$}\nointerlineskip\hbox{$\sim$}}}}
\def\gsim{\mathrel{\vcenter{\hbox{$>$}\nointerlineskip\hbox{$\sim$}}}}

\begin{document}
\thispagestyle{empty}
\begin{flushright} UCRHEP-T279\\June 2000\
\end{flushright}
\vspace{0.5in}
\begin{center}
{\Large \bf Verifiable Model of Neutrino Masses\\
from Large Extra Dimensions\\}
\vspace{1.0in}
{\bf Ernest Ma$^1$, Martti Raidal$^{1,2}$, and Utpal Sarkar$^{1,3}$\\}
\vspace{0.2in}
{$^1$ \sl Department of Physics, University of California, Riverside, 
California 92521, USA\\}
\vspace{0.1in}
{$^2$ \sl National Institute of Chemical and Biological Physics, R\"avala 10, 
10143, Estonia\\}
\vspace{0.1in}
{$^3$ \sl Physical Research Laboratory, Ahmedabad 380 009, India\\}
\vspace{1.0in}
\end{center}
\begin{abstract}

We propose a new scenario of neutrino masses with a Higgs triplet $(\xi^{++}, 
\xi^+, \xi^0)$ in a theory of large extra dimensions.  Lepton number violation 
in a distant brane acts as the source of a very small trilinear coupling of 
$\xi$ to the standard Higgs doublet in our brane.  Small realistic Majorana 
neutrino masses are \underline{naturally} obtained with the fundamental scale 
$M_* \sim {\cal O}(1)$ TeV, foretelling the possible discovery of $\xi$ 
$(m_\xi\lsim M_*)$ at future colliders.   Decays of $\xi^{++}$ into same-sign 
dileptons are fixed by the neutrino mass matrix.  Observation of $\mu-e$ 
conversion in nuclei is predicted.

\end{abstract}

\newpage
\baselineskip 24pt

Recently it has been proposed \cite{extra1} that the fundamental scale of 
quantum gravity could be as low as a few TeV.  In this theory, our 
four-dimensional world (called a 3-brane) is localized to only one point in 
a space having $n$ extra dimensions (the bulk).  Assuming that gravity is 
the only field which propagates in these extra dimensions, the fact that it 
is so weak in our world may be understood.  Since there are no scales larger 
than a few TeV in this theory, the smallness of neutrino masses requires 
a different explanation than the usual one that it is inversely proportional 
to some very large mass.  There are already proposals in this direction, 
where singlet fermions are introduced in the bulk \cite{exnu1,exnu2}.  They 
become the right-handed partners of the observed left-handed neutrinos, but 
with small Dirac masses -- the result of being suppressed by the volume of 
the extra dimensions.  Note first that this mechanism by itself does not 
generate a Majorana mass.  Note also that such a model of neutrino masses, 
like any other before it, has no distinct experimental signatures which would 
set it apart from any other.  Furthermore, the existence of singlet fermions 
in the bulk is strongly constrained by supernova data to have an upper bound 
of $210/\sqrt{n}$ TeV on the fundamental scale involved \cite{exsn}. 

In this Letter we propose a simple, verifiable scenario in which the distant 
breaking \cite{distant} of lepton number is used to generate small Majorana 
neutrino masses through a scalar Higgs triplet \cite{trip} localized in our 
brane.  The trilinear interaction of the triplet with the standard Higgs 
doublet is induced by the ``shining'' of a scalar singlet which communicates 
the violation of lepton number from a distant brane to our world through 
the large extra space dimensions.  Thus the amount of lepton number violation 
in our world is suppressed by the distance between us and the other brane. 
Because the Higgs triplet should not be heavier than the fundamental scale 
of the extra dimensions, this scenario provides the basis for a rich and 
robust phenomenology of new observable processes at future high-energy 
colliders as well as low-energy precision experiments.

The Higgs triplet may be produced at the upgraded Tevatron at Fermilab, the 
Large Hadron Collider (LHC) at CERN, and perhaps a future linear collider 
(LC) or muon collider (MC) or both.  The decay 
branching fractions of the doubly charged member $\xi^{++}$ of the Higgs 
triplet into same-sign dileptons or the cross sections of the resonant 
processes $e^-e^- (\mu^-\mu^-) \to l^-_i l^-_j$ may then be used to 
\underline {determine} the magnitude of \underline {each} element of the 
neutrino mass matrix completely and uniquely up to an overall scale factor.  
Whereas lepton-number violating processes are suppressed at low energies 
(because their amplitudes are proportional to the neutrino mass matrix), large 
lepton-flavor violation through the Higgs triplet is possible.  Based on 
current neutrino-oscillation data, the rate of coherent $\mu-e$ conversion 
in nuclei, with an amplitude enhanced by the large factor 
$\ln(m^2_{\xi^{++}}/m_\mu^2)$ over that of $\mu \to e \gamma$, is predicted 
to be observable at the proposed Muon Electron Conversion (MECO) experiment 
\cite{meco} at Brookhaven National Laboratory (BNL).

We extend the minimal standard model of particle interactions to include 
a heavy Higgs triplet $(\xi^{++}, \xi^+, \xi^0)$ with the interaction
\begin{equation}
{\cal L} = f_{ij} [ \xi^0 \nu_i \nu_j + 
\xi^+\frac{(\nu_i l_j + l_i \nu_j)}{\sqrt{2}} + \xi^{++} l_i l_j] + h.c.,
\end{equation}
implying that $\xi$ carries lepton number $L = -2$.  The resulting Majorana 
neutrino mass matrix is then given by
\begin{equation}
({\cal M}_\nu)_{ij} = 2 f_{ij} \langle \xi^0 \rangle,
\end{equation}
where $\langle \xi^0 \rangle$ is the vacuum expectation value (VEV) of 
$\xi^0$, hence lepton number must be violated somehow.  This may be achieved 
with the following addition to the Lagrangian:
\begin{equation}
{\cal L}' =
h \chi (\bar \xi^0 \phi^0 \phi^0 - \sqrt 2 \xi^- \phi^+ \phi^0 
+ \xi^{--} \phi^+ \phi^+) + h.c.,
\end{equation}
where $\chi$ is a complex neutral scalar singlet with $L = -2$.  We then 
obtain \cite{trip}
\begin{equation}
\langle \xi^0 \rangle \simeq {h \langle \chi \rangle \langle \phi^0 \rangle^2 
\over m_\xi^2}.
\end{equation}
In other words, the breaking of $L$ by $\langle \chi \rangle$ is communicated 
to the lepton sector through $\langle \xi \rangle$.  As the neutrino masses 
are very small, we need $m_\xi$ to be very large in Eq.~(4).  Suppose $m_\xi$ 
is only of order 1 TeV, then $h \langle \chi \rangle$ would have to be very 
small, i.e. $h \langle \chi \rangle \sim 33$ eV, if $\langle \xi^0 \rangle 
\sim 1$ eV, using $\langle \phi^0 \rangle = 174$ GeV.  However, $\chi$ is a 
singlet, and as such, its VEV is expected to be large, so this requires what 
we normally would consider as extremely unnatural fine tuning.  However, in a 
theory of large extra dimensions, a small $\langle \chi \rangle$ may arise 
\underline{naturally}, thus allowing $m_\xi$ to be of order 1 TeV and be 
directly observed experimentally.

The standard model fields together with $\xi$ are localized in our world 
(a 3-brane ${\cal P}$ at $y = 0$) and are blind to the extra space dimensions.
Lepton number is assumed to be conserved as far as these fields are concerned. 
The singlet $\chi$ is special; it propagates also in the bulk carrying lepton 
number $L = -2$ and interacts in our brane according to Eq.~(3).  The 
violation of lepton number happens at a distant 3-brane which then gets 
communicated to our brane through $\langle \chi \rangle$.

We assume the existence of a field $\eta$ which is localized in a distant 
3-brane (${\cal P}'$) situated at a point $y = y_*$ in the extra dimensions. 
It is a singlet under the standard model but has $L = 2$ and couples to 
$\chi$ (with $L=-2$).  When the field $\eta$ acquires a VEV, lepton number is 
broken maximally in the other brane.  It will then act as a point source for 
$L$ violation, and the field $\chi$ is the messenger which carries it to our 
wall (the interface between our brane and the bulk).  The ``shining'' of 
$\chi$ at all points in our world is the mechanism \cite{distant} which 
breaks lepton number and gives mass to the neutrinos.

At energies much below the fundamental scale $M_*$, the lepton-number 
violating effect will be suppressed by the distance between the source brane 
at ${\cal P}'$ and our brane at ${\cal P}$.  We assume that the source brane 
is situated at the farthest point in the extra dimensions so that $|y_*| = r$ 
is the radius of compactification and it is related to the fundamental scale 
$M_*$ and the reduced Planck scale ($M_P=2.4 \times 10^{18}$ GeV) by the 
relation
\begin{equation}
r^n M_*^{n + 2} \sim M_P^2  .
\end{equation}
This explains why lepton number is only violated weakly in our world.

We assume here that the source brane has the same dimensional structure as 
our world and there are $n$ extra dimensions.  In our world (${\cal P}$) the 
field $\chi$ has only the lepton-number conserving interaction
\begin{equation}
{\cal S}_{\chi} = h ~\int_{\cal P} d^4 x ~ \xi^\dagger(x) \phi(x) \phi (x) 
\chi(x, y=0).
\end{equation}
In the other brane (${\cal P}'$) the field $\chi$ couples to the field 
$\eta$ through the interaction
\begin{equation}
{\cal S}_{other} = \int_{{\cal P}'} d^4 x' ~\mu^2~ \eta (x') \chi 
(x', y = y_*), 
\end{equation}
where $\mu$ is a mass parameter.  Lepton number violation from $\langle \eta 
\rangle$ is carried by $\chi$ to our world through its ``shined'' value 
$\langle \chi \rangle$: 
\begin{equation}
\langle \chi (x, y = 0) \rangle = \Delta_n (r) \langle \eta (x, y = y_*) 
\rangle,
\end{equation}
where $\langle \eta \rangle$ acts as a point source, and $\Delta_n(r)$ is 
the Yukawa potential in $n$ transverse dimensions, i.e. \cite{distant}
\begin{equation}
\Delta_n (r)= {1 \over (2 \pi )^{n \over 2}
M_*^{n- 2}} ~\left( {m_\chi \over r} \right)^{n-2 \over 2}
~K_{n - 2 \over 2} \left( m_\chi r \right),
\end{equation}
$K$ being the modified Bessel function.  If the mass of the carrier 
field $\chi$ is large ($m_\chi r \gg 1$), it has the profile
\bea
\langle \chi \rangle \approx \displaystyle{ 
m_\chi^{n - 3 \over 2} \over 2 (2 \pi)^{n - 1 
\over 2} M_*^{ n - 3} } \displaystyle{e^{- 
m_\chi r} \over r^{n-1 \over 2} }.
\eea
The suppression here is exponential, hence the amount of lepton-number 
violation in our world is very small, but its precise value depends 
sensitively on $m_\chi$.  An interesting alternative is to have a light 
carrier field with a mass less than $1/r$.  However, it should be larger 
than about (0.1 mm)$^{-1}$, to be consistent with the present experimental 
data on gravitational interactions.

If $m_\chi r \ll 1$, $\Delta_n(r)$ is logarithmic for $n=2$ and $\langle 
\chi \rangle$ is not suppressed.  For $n>2$, the asymptotic form of the 
profile of $\chi$ is
\begin{equation}
\langle \chi \rangle \approx { \Gamma ( {n -2 \over 2} ) \over
4 \pi^{n \over 2} }{M_* \over (M_* r)^{n-2} },
\end{equation}
which is suitably small for large $r$.  Using Eqs.~(2), (4), and (5), we 
then obtain
\begin{equation}
({\cal M}_\nu)_{ij}  \approx {\Gamma ({n-2 \over 2})
\over 2 \pi^{n \over 2}} ~ h~ f_{ij}~ {\langle \phi \rangle^2 ~ M_*
\over m_\xi^2 } \left( {M_* \over M_P} \right)^{2 - {4 \over n}} .
\label{mn}
\end{equation}
For $n =3$,  $M_* \sim 1$ TeV,  $m_\xi \lsim 1$ TeV and $h \sim f_{33}\sim 
0.25$, we get $({\cal M}_\nu)_{33} \sim 0.03$ eV, which is of the right 
magnitude for understanding atmospheric neutrino oscillations.  There is a 
massless Goldstone boson (Majoron) corresponding to the spontaneous breaking 
of lepton number in this scenario.  It is a singlet and its coupling to 
the leptons in our world is suppressed by $\langle \eta \rangle^{-1}$.  It 
is thus completely acceptable phenomenologically. 

For $n \geq 4$, this scenario requires a larger fundamental scale $M_*$ and 
is of less phenomenological interest.  However, we have assumed here for 
simplicity the dimension of the source brane to be the same as ours, which 
may not be the case.  The distance between the branes may also be smaller.  
On the other hand, such complications diminish the predictive power of the 
proposed mechanism; they will not be considered any further.

\begin{figure}[t]
\centerline{
\epsfxsize = 0.5\textwidth \epsffile{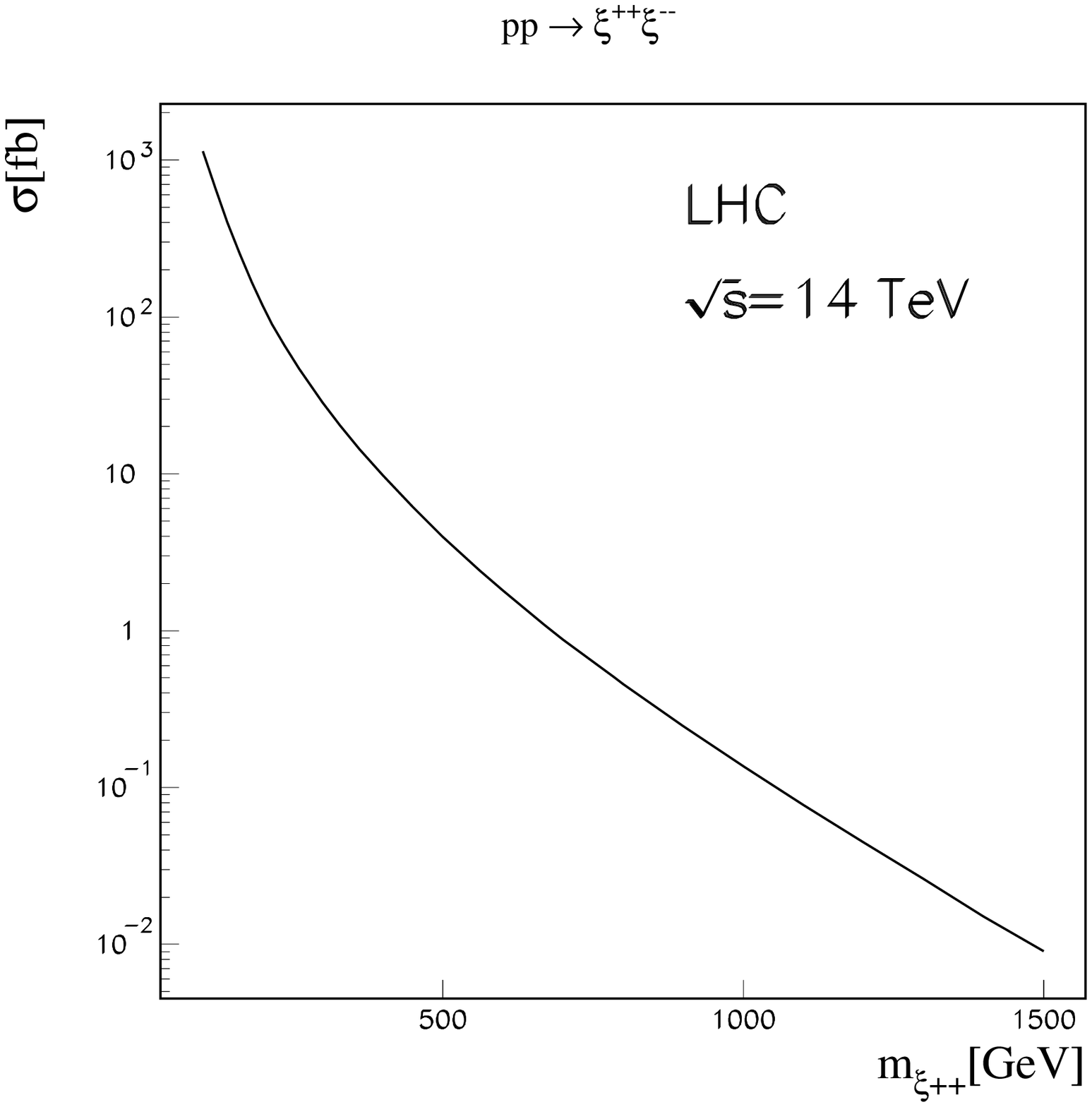}
}
\caption{Cross section of $\xi^{++}\xi^{--}$ Drell-Yan pair production 
at LHC. }
\label{fig:ppdd}
\end{figure}

The only pair production mechanism of $\xi^{++}$ at the LHC is the Drell-Yan 
process mediated by s-channel photon and Z-boson exchange \cite{hmpr}. 
Thus the production rate is enhanced by the double charge of $\xi^{++}$ 
and is uniquely determined by the gauge couplings.  To obtain the cross 
section of $pp\to\xi^{++}\xi^{--}$ at the LHC, we have calculated the 
subprocesses involving the $u,\,\bar u$ and $d,\,\bar d$ collisions and 
convoluted them over the parton distributions given by the default set of the 
CERN library package PDFLIB \cite{pdflib}.  The total cross section 
as a function of the triplet mass $m_{\xi^{++}}$ is plotted in 
Fig.~\ref{fig:ppdd}.

Once produced, $\xi^{++}$ will decay into same-sign dileptons.  Because  
$\langle \xi\rangle$ is tiny, the decay branching fraction of $\xi^{++}\to 
W^+W^+$ is negligible.  Thus the \underline {only} possible decay channels 
in our scenario are $\xi^{++}\to l_i^+l_j^+$ with the partial rates 
$\Gamma_{ij}=|f_{ij}|^2 m_\xi/4\pi$ for $i\neq j,$ and 1/2 smaller for $i=j.$  
This same-sign dilepton signal at the invariant mass of $\xi$ is very 
distinctive at the LHC because it is completely \underline {background-free}. 
Assuming the total integrated luminosity of the LHC to be 1000 $fb^{-1},$ 
the reconstruction efficiency of the event to be 10\% (to be very conservative)
and that the predicted average of $N=-\ln(1-p)$ Poisson distributed 
events would provide a discovery, the cross section in Fig.~\ref{fig:ppdd} 
implies at $p=95\%$ confidence level that $m_{\xi^{++}}\lsim 1.2$ TeV can be 
probed at the LHC [300 GeV at the Tevatron].  Its decay branching fractions 
will then determine $|f_{ij}|$, i.e. the magnitude of each element of the 
neutrino mass matrix up to an overall scale factor.  This is the only model 
of neutrino masses which has the promise of being verified from collider 
experiments.

Complementary measurement of $|f_{ij}|$ is provided by the resonant 
processes $e^-e^-(\mu^-\mu^-)\to l_i^-l_j^-$ at a future LC and/or MC. 
The $\xi^{++}$ mass reach in these colliders extends up to the collision 
energies, which may be as high as 4 TeV.  The sensitivity to $|f_{ij}|$ 
depends on the beam properties of the machines.  The detailed estimate 
in Ref.~\cite{cr} implies that $|(f\cdot f^*)_{ij}|\gsim 10^{-8}$ can be
probed in these processes.   

Consider now a phenomenological hierarchical neutrino mass matrix consistent 
with the atmospheric and solar neutrino results \cite{neutrino}:
\bea
{\cal M}_\nu  = m \pmatrix{ a x & b & b x \cr b & x^2 & x \cr b x
& x & 1} ,
\label{mnmatr}
\eea
where $m$ is the normalization mass and $0.67 < x < 1$ determines the 
$\nu_\mu \to \nu_\tau$ mixing as required by the atmospheric neutrinos.  
The three solutions of the solar neutrino problem correspond to (i) 
large-angle matter-enhanced oscillations: $a=0.006,~b=0.04$; (ii) small-angle 
matter-enhanced oscillations: $a=4 \times 10^{-6};~b=0.0012$; and (iii) 
vacuum oscillations: $a = 0.0002,~b=0.0012$.

Given the pattern of $f_{ij}$ via the neutrino mass matrix of 
Eq.~\rfn{mnmatr}, lepton-flavor violation through $\xi$ exchange may be 
observable at low energies.  The amplitude of $\mu-e$ conversion in nuclei 
is enhanced by $\ln(m^2_{\xi^{++}}/m_\mu^2)$ compared to that of $\mu\to e
\gamma$ \cite{rs}.  Planned experiments will reach the sensitivity of 
$10^{-16}$ for $\mu-e$ conversion in aluminum \cite{meco} and $10^{-14}$ for 
$\mu\to e\gamma$ \cite{mueg}. The matrix element of photonic conversion 
is given by ${\cal M}=(4\pi\alpha/q^2) j^{\mu} J_{\mu}$, where $q$ is the 
momentum transfer with $q^2 \approx -m^2_\mu,$ $J$ is the hadronic current, 
and
\bea
j^{\lambda} &=& {\bar u}(p_e) \left[ \; 
\left (f_{E0} + \gamma_5 f_{M0}\right) \gamma_{\nu}  \left (
g^{\lambda \nu} - \frac {q^{\lambda} q^{\nu}}{q^2}\right )
\right. \nn \\
&  & +
\left. 
(f_{M1} + \gamma_5 f_{E1})\; i\; \sigma^{\lambda \nu} \frac{q_{\nu}}{m_{\mu}} 
\right] u(p_{\mu})\,
\label{j1}
\eea
is the leptonic current.  The coherent $\mu-e$ conversion ratio in nuclei 
is given by
\beq
R_{\mu e}=\frac{8\alpha^5\,m_\mu^5\,Z^4_{eff}\,Z\,|\overline{F_p}(p_e)|^2}
{\Gamma_{capt}}
 \frac{\xi_0^2}{q^4}\, ,
\label{mecrate}
\eeq
where $\xi_0^2=|f_{E0}+f_{M1}|^2+|f_{E1}+f_{M0}|^2$, and for $^{13}$Al, 
$Z^{Al}_{eff}=11.62,$  $\overline{F_p}^{Al}(q)=0.66$, and
$\Gamma_{capt}^{Al}=7.1\times 10^5$~s$^{-1}$ \cite{chiang}.
We calculate the form factors induced by the one-loop diagrams involving
$\xi^{++}$ and obtain  
\bea
&& f_{E0} = f_{M0}= \sum_l {f_{\mu l} f^*_{l e} \over 24\pi^2} [4s_l+rF(s_l)], 
\\
&& f_{M1} = -f_{E1}= \sum_l {f_{\mu l} f^*_{l e} \over 24\pi^2} s_\mu,
\eea
where $$F(s_l) =
\ln s_l+\left(1-\frac{2s_l}{r}\right)\sqrt{1+\frac{4s_l}{r}}
\ln\left[\frac{\sqrt{r+4s_l}+\sqrt{r}}{\sqrt{r+4s_l}-\sqrt{r}}\right],
$$
and $r=-q^2/m_{\xi^{++}}^2$, $s_l=m_l^2/m_{\xi^{++}}^2,$ $l=e,\,\mu,\,\tau$. 
In the interesting limit $s_l\to 0$, we get $F(s_l)\to \ln r.$

\begin{figure}[t]
\centerline{
\epsfxsize = 0.5\textwidth \epsffile{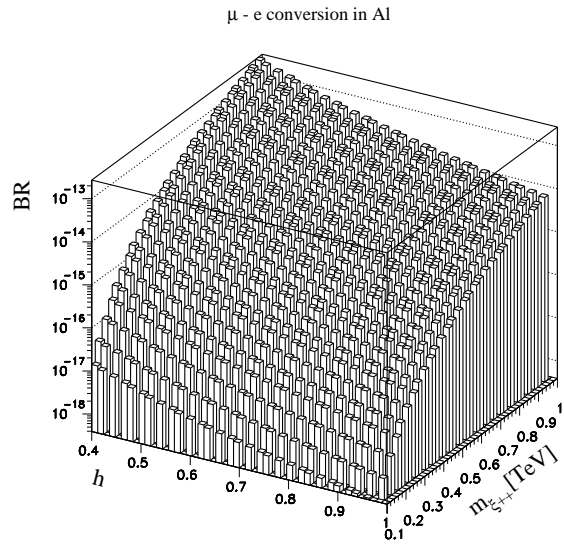}
}
\caption{Rate of $\mu-e$ conversion in $^{13}$Al against the $\xi^{++}$ mass
and Higgs self coupling $h$ for $M_*=1$ TeV, assuming large-angle 
matter-enhanced solution to the solar neutrino problem.  }
\label{fig:muec}
\end{figure}

For numerical estimates we assume $m=0.03$ eV in Eq.~(13) and the large-angle 
matter-enhanced oscillation solution to the solar neutrino problem. 
In Fig.~\ref{fig:muec} we plot the ratio of $\mu-e$ conversion in 
aluminum as a function of the mass $m_{\xi^{++}}$ and coupling $h$, 
assuming the fundamental scale to be $M_*=1$ TeV.  The behavior of this 
ratio can be understood from \Eq{mn}: a fixed neutrino mass implies 
$f \propto m_{\xi^{++}}^2 $ and $f \propto 1/h.$  Notice the complementarity 
of collider and $\mu-e$ conversion experiments.  For small $m_{\xi^{++}}$, 
$R_{\mu e}$ is suppressed while the collider cross section is kinematically 
enhanced, and vice versa.   Thus for $M_*= 1$ TeV, the MECO experiment 
should see a signal unless $m_{\xi^{++}}\lsim 300$ GeV.  If $m_{\xi^{++}}
\sim M_*$, which is the most likely situation in our scenario, and $h\sim 1$ 
which minimizes the signal, then MECO will test our model up to the scale 
$M_*\sim 7$ TeV.

In conclusion, we have proposed a verifiable scenario of large extra 
dimensions in which we invoke the distant breaking of lepton number to 
obtain small Majorana neutrino masses, using a Higgs triplet $\xi$ localized 
in our world.  In this scenario, the fundamental scale $M_*$ may be a few 
TeV, thus $m_\xi \lsim M_*$ makes $\xi$ kinematically accessible at 
the LHC and at future lepton colliders.  Our model has the unique feature 
that the decay branching fractions of $\xi^{++}$ into charged leptons $l^+_i 
l^+_j$ determine directly $|f_{ij}|$, where $f_{ij}$ is the neutrino mass 
matrix up to an overall scale factor.  Using present neutrino data, we also 
predict observable $\mu-e$ conversion in nuclei.  In particular, for 
hierarchical neutrino masses and the large-angle matter-enhanced solution 
for the solar neutrino problem, the MECO experiment can test our model up to 
$m_\xi\sim M_*\sim 7$ TeV. 

{\it Acknowledgement.} This work was supported in part by the U.~S.~Department 
of Energy under Grant No.~DE-FG03-94ER40837. 

\bibliographystyle{unsrt}


\end{document}